\documentclass[epjST]{svjour}
\usepackage{graphics}
\usepackage{cite}

\newcommand{\be}{\begin{equation}}
\newcommand{\ee}{\end{equation}}
\newcommand{\ba}{\begin{eqnarray}}
\newcommand{\ea}{\end{eqnarray}}
\newcommand{\la}{\langle}
\newcommand{\ra}{\rangle}
\newcommand{\di}{ {\rm d} }
\newcommand{\ph}{$\phantom{1}$}

\begin{document}
\title{TMDs in the bag model\thanks{Also reported at 
Spin Symposium 2010, Juelich, Germany.}}
\author{H Avakian\inst{1} \and 
\underline{A V Efremov}\inst{2}\fnmsep\thanks{\email{efremov@theor.jinr.ru}} \and P Schweitzer\inst{3} }
\institute{Thomas Jefferson National Accelerator Facility,
Newport News, VA 23606, U.S.A. \and 
Joint Institute for Nuclear Research, Dubna,
141980 Russia \and Department of Physics, University of Connecticut,
Storrs, CT 06269, U.S.A.}
\abstract{
Leading and subleading twist transverse momentum dependent parton 
distribution functions  (TMDs) are studied in a quark model 
framework provided by the bag model. A complete set of relations 
among different TMDs is derived, and the question is discussed 
how model-(in)dependent such relations are. A connection of the 
pretzelosity distribution and quark orbital angular momentum is 
derived.
} 
\maketitle
\section{Introduction}
TMDs are a generalization 
of parton distribution functions (PDFs) promising to extend our 
knowledge of the nucleon structure far beyond what we have 
learned from PDFs about the longitudinal momentum distributions 
of partons in the nucleon. In addition to the latter, TMDs carry 
also information on transverse parton momenta and spin-orbit 
correlations. TMDs enter the description of leading-twist 
observables in deeply inelastic reactions 
on which data are 
available like: semi-inclusive deep-inelastic scattering (SIDIS),
Drell-Yan process 
or hadron production in $e^+e^-$ annihilations. 
The interpretation of these data is not straight-forward.
In SIDIS one deals with convolutions of a priori unknown 
transverse momentum distributions in nucleon and fragmentation 
process, and in practice is forced to {\sl assume} models 
such as the Gaussian Ansatz. 
In the case of subleading twist observables, one moreover faces 
the problem that several twist-3 TMDs and fragmentation functions 
enter the description of one observable 
(we recall that presently factorization is not proven for 
subleading-twist observables).

In this situation information from models is valuable for several 
reasons. Models can be used for direct estimates of observables. 
An equally interesting aspect concerns relations among different 
TMDs observed in models. 
Such relations, especially when supported by several models, 
could be helpful --- at least for qualitative interpretations of 
first data. Model results also allow to test assumptions  
made in literature, such as the Gaussian Ansatz for transverse 
momentum distributions or certain approximations. 
In addition to that model studies are of interest also because they 
provide important insights into non-perturbative properties of TMDs. 

The purpose of this talk is to review the main results for 
relations among TMDs derived in the MIT bag model. In this model, 
quark-quark correlation functions in the nucleon 
\cite{Bacchetta:2006tn} can be expressed in terms of a quark 
wave-functions, which have an $S$-wave component proportional to 
the function $t_0(k)$ and a $P$-wave component proportional to 
$t_1(k)$, where $k=|\vec{k}|$ is the quark momentum. The presence 
of these components is a minimal requirement for the modelling of 
T-even TMDs. Since there are no explicit gluon degrees of 
freedom, T-odd TMDs vanish in this model. We restrict ourselves 
to T-even TMDs. More details and references can be found in  
\cite{Avakian:2010br}. 

\section{TMDs in the bag model}
\label{Sec-2:TMDs-in-bag}

We assume $SU(6)$ spin-flavor symmetry of the proton wave 
function. In SU(6) spin-independent TMDs of definite flavor are 
given in terms of 'flavor-less' expressions multiplied by a 
'flavor factor' $N_q$ with $N_u = 2$, $N_d=1$. Spin-dependent 
TMDs of definite flavor follow from multiplying 'flavor-less' 
expressions by a 'spin-flavor factor' $P_q$ with $P_u = 
\frac{4}{3}$, $P_d = -\frac{1}{3}$.

Defining $\hat{k} = \vec{k}/k$ and $\widehat{M}_M = M_N/k$, 
the results for T-even leading twist TMDs read
\ba
f_1^q(x,k_\perp)=N_qA\bigl[t_0^2+2\widehat{k}_z\,t_0t_1+t_1^2\bigr],&&
g_1^q(x,k_\perp) = P_q\,A\bigl[t_0^2+2\widehat{k}_z\,t_0t_1
+(2\widehat{k}_z^2-1)\,t_1^2\bigr], 
\nonumber\\ 
h_1^q(x,k_\perp) = P_q\,A\bigl[t_0^2+2\widehat{k}_z\,t_0t_1
+\widehat{k}_z^2\,t_1^2\bigr],      &&
g_{1T}^{\perp q}(x,k_\perp) = P_q\,A\bigl[\phantom{-}2\widehat{M}_N
(t_0t_1+\widehat{k}_z\,t_1^2)\bigr],
\nonumber\\ 
h_{1L}^{\perp q}(x,k_\perp) = P_q\,A\bigl[-2\widehat{M}_N
(t_0t_1+\widehat{k}_z\,t_1^2)\bigr], &&
h_{1T}^{\perp q}(x,k_\perp)=P_q\,
A\bigl[-2\widehat{M}_N^{\,2}\,t_1^2\bigr],
\nonumber 
\ea
and for the subleading twist TMDs we obtain
\ba
e^q(x,k_\perp)=N_q  A\bigl[t_0^2-t_1^2 \bigr], && 
f^{\perp q}(x,k_\perp)=N_qA\bigl[2\widehat{M}_N\,t_0t_1\bigr],
\nonumber\\
g_T^q(x,k_\perp)=P_q\,A\bigl[t_0^2-\widehat{k}_z^2\,t_1^2\bigr],&&
g_L^{\perp q}(x,k_\perp)=P_q\,A\bigl[2\widehat{M}_N\,\widehat{k}_z\,t_1^2\bigr],
\nonumber\\ 
g_T^{\perp q}(x,k_\perp)=P_q\,A\bigl[2\widehat{M}_N^2\,t_1^2\bigr],&& 
h_L^q(x,k_\perp)=P_q\,A\bigl[t_0^2+(1-2\widehat{k}_z^2)t_1^2\,\bigr],
\nonumber\\ 
h_T^{\perp q}(x,k_\perp)=P_q\,A\bigl[2\widehat{M}_N\;t_0t_1\bigr],&& 
h_T^{q}(x,k_\perp)=P_q\,A\bigl[-2\widehat{M}_N\widehat{k}_z\,t_1^2\bigr] ,
\nonumber 
\ea   
where $A$ is a common normalization factor \cite{Avakian:2010br}. 

\section{Relations among TMDs in bag model}
In the bag model, there are 9 linear relations among the 14 
(twist-2 and 3) T-even TMDs, which can be written as follows 
(where $j^{(1)q}(x,k_\perp)= 
\frac{k_\perp^2}{2M_N^2}j^q(x,k_\perp)$ and the 'dilution 
factor'  ${\cal D}^q=\frac{P_q}{N_q}$), 

{\small 
\vspace{-4mm}%
\hskip-7mm%
\begin{minipage}[t]{0.5\textwidth}
\be{\cal D}^q\,f_1^q(x,k_\perp)+g_1^q(x,k_\perp)= 
2h_1^q(x,k_\perp)\label{Eq:rel-I},\ee\\[-6.5mm]%
\be {\cal D}^q\,e^q(x,k_\perp)+h_L^q(x,k_\perp)=2g_T^q(x,k_\perp)
\label{Eq:rel-II},\ee\\[-5mm]%
\be{\cal D}^q\,f^{\perp q}(x,k_\perp)=h_T^{\perp q}(x,k_\perp)
\label{Eq:rel-III},\ee\\[-5mm]%
\be g_{1T}^{\perp q}(x,k_\perp)=-h_{1L}^{\perp q}(x,k_\perp),
\label{Eq:rel-IV}\ee,\\[-6mm]%
\be g_T^{\perp q}(x,k_\perp)=-h_{1T}^{\perp q}(x,k_\perp) 
\label{Eq:rel-V},\ee
\end{minipage}
\hfil
\begin{minipage}[t]{0.5\textwidth}
\vskip3mm%
\be g_L^{\perp q}(x,k_\perp)=-h_T^{q}(x,k_\perp)\label{Eq:rel-VI},\ee\\[-5mm]%
\be g_1^q(x,k_\perp)-h_1^q(x,k_\perp)=h_{1T}^{\perp(1)q}(x,k_\perp)
\label{Eq:measure-of-relativity},\ee\\[-5mm]%
\be g_T^{q}(x,k_\perp)-h_L^{q}(x,k_\perp)= 
h_{1T}^{\perp(1)q}(x,k_\perp)\label{Eq:rel-VIII},\ee\\[-5mm]%
\be h_T^{q}(x,k_\perp)-h_T^{\perp q}(x,k_\perp)=h_{1L}^{\perp 
q}(x,k_\perp) \label{Eq:rel-IX}\,.\ee\\[-5mm]%
\ph\hfill 
\end{minipage}
} 
\vspace{2mm}%

Why are there 9 linear relations? In this context, the 5 structures
$t_0^2$, $t_0t_1$, $t_1^2$, $\widehat{k}_zt_0t_1$, $\widehat{k}_zt_1$ 
are to be considered as linearly independent. This implies 
9 linear equations among the 14 TMDs.

However, there are also {\sl non-linear relations}, for example, 
\ba
\label{Eq:non-lin-2}
h_1^q(x,k_\perp)\,h_{1T}^{\perp q}(x,k_\perp) &=& 
-\frac{1}{2}\,\biggl[h_{1L}^{\perp q}(x,k_\perp)\biggr]^2,\\ 
\label{Eq:non-lin-1}
g_T^q(x,k_\perp)\,g_T^{\perp q}(x,k_\perp) &=& 
\phantom{-}\frac{1}{2}\,\biggl[g_{1T}^{\perp q}(x,k_\perp)\biggr]^2 -
g_{1T}^{\perp q}(x,k_\perp)\,g_L^{\perp q}(x,k_\perp)\,.
\ea 
The Eqs.~(\ref{Eq:non-lin-2},~\ref{Eq:non-lin-1}) are independent 
in the sense that it is impossible to convert one into the other 
upon use of the linear relations 
(\ref{Eq:rel-I}--\ref{Eq:rel-IX}). With the 9 linear relations 
(\ref{Eq:rel-I}--\ref{Eq:rel-IX}) and the 2 non-linear relations 
(\ref{Eq:non-lin-2},~\ref{Eq:non-lin-1}) we find altogether 11 
relations among 14 TMDs in the bag model.      

The deeper reason, why  in the bag model relations among TMDs 
appear, is ultimately related to Melosh rotations which connect 
longitudinal and transverse nucleon and quark polarization states 
in a Lorentz-invariant way \cite{Efremov:2002qh}. This was 
elucidated in detail in Ref.\cite{Lorce:2011zt}. 

\subsection{Comparison to other quark models} 
An important issue, when observing relations among TMDs in a 
model, concerns their presumed validity beyond that particular 
model framework. For that it is instructive to compare to other 
models.
\begin{itemize}
\item 
Eq.~(\ref{Eq:rel-I}): its  $k_\perp$-integrated version was 
discussed in bag model in \cite{Jaffe:1991ra} and 
\cite{Signal:1997ct,Barone:2001sp} and in light-cone constituent 
models in \cite{Pasquini:2005dk}. The unintegrated version was 
discussed in bag and light-cone constituent models 
\cite{Pasquini:2008ax,Avakian:2008dz}.
\item 
Eq.~(\ref{Eq:rel-II}): its integrated version was observed in the 
bag model previously in \cite{Signal:1997ct}.
\item 
Eq.~(\ref{Eq:rel-IV}): was first observed in the spectator model 
of \cite{Jakob:1997wg} and later also in light-cone constituent 
models \cite{Pasquini:2008ax} and the covariant parton model of 
Ref.~\cite{Efremov:2009ze}.
\item 
Eq.~(\ref{Eq:rel-VI}): 
was found in the spectator model of Ref.~\cite{Jakob:1997wg}.
\item 
Eq.~(\ref{Eq:measure-of-relativity}): was first observed in the 
bag \cite{Avakian:2008dz}. It is valid also in the spectator 
\cite{Jakob:1997wg}, light-cone constituent 
\cite{Pasquini:2008ax}, and covariant parton 
\cite{Efremov:2009ze} models.
\item
Eqs.~(\ref{Eq:rel-III},~\ref{Eq:rel-V},~\ref{Eq:rel-VIII},~\ref{Eq:rel-IX}):
are new in the sense of not having been mentioned previously in 
literature. But the latter 3 are satisfied by the spectator model 
results from \cite{Jakob:1997wg}.
\item 
The non-linear relation (\ref{Eq:non-lin-2}) connecting all 
T-even, chiral-odd leading-twist TMDs was found in the covariant 
parton model approach \cite{Efremov:2009ze}. 
Eq.~(\ref{Eq:non-lin-1}) was not discussed prior to 
\cite{Avakian:2010br}.
\end{itemize}
The detailed comparison, in which models these relations hold and 
in which they are violated, gives some insight into the question 
to which extent these relations are model-dependent.

Let us discuss first Eqs.~(\ref{Eq:rel-I}--\ref{Eq:rel-III}), 
which connect polarized and unpolarized TMDs. For these relations 
SU(6)-spin-flavor symmetry is necessary, but not sufficient. For 
example, the spectator model of \cite{Jakob:1997wg} is SU(6) 
symmetric. But it does not support 
(\ref{Eq:rel-I}--\ref{Eq:rel-III}) which are spoiled by the 
different masses of the (scalar and axial-vector) spectator 
diquark systems. Also (\ref{Eq:rel-I},~\ref{Eq:rel-II}) are not 
supported in the covariant parton model approach of 
\cite{Efremov:2009ze}. However, in that approach it is 
possible to 'restore' these relations by introducing additional, 
restrictive assumptions, see \cite{Efremov:2009ze} for a detailed 
discussion. We conclude that the relations 
(\ref{Eq:rel-I}--\ref{Eq:rel-III}) require strong model 
assumptions. It is difficult to estimate to which extent such 
relations could be useful approximations in nature, though they 
could hold in the valence-$x$ region with an accuracy of 
(20--30)$\,\%$ (see \cite{Boffi:2009sh}).

From the point of view of model dependence, it is 'safer' 
\cite{Avakian:2008dz} to compare relations which include only 
polarized TMDs such as Eqs.~(\ref{Eq:rel-IV}--\ref{Eq:rel-IX}).
(Or, only unpolarized TMDs, for which we know no example.)
It is gratifying to observe that these 
relations are satisfied not only in the bag model, but also in 
the spectator model version of Ref.~\cite{Jakob:1997wg}. The 
relations among the leading twist TMDs, 
Eqs.~(\ref{Eq:rel-IV},~\ref{Eq:measure-of-relativity}), hold also 
in light-cone constituent \cite{Pasquini:2008ax}, and covariant 
parton \cite{Efremov:2009ze} models. 

Of course,  quark model relations among TMDs have limitations, 
even in quark models. In \cite{Bacchetta:2008af} various versions 
of spectator models were used, and in some versions the relations 
were not supported 
(\ref{Eq:rel-IV},~\ref{Eq:measure-of-relativity}). Also the 
quark-target model \cite{Meissner:2007rx} does not support 
(\ref{Eq:rel-IV},~\ref{Eq:measure-of-relativity}). 
Finally, in QCD none of such relations is valid, and all TMDs are 
independent structures. It would be interesting to `test' the
quark model relations in other models, lattice QCD, and in 
experiment. 

\subsection{A relation among PDFs} It is worth to discuss in some 
more detail one particularly interesting relation, which includes 
only functions known from the collinear case. By eliminating the 
transverse moment of the pretzelosity distribution from 
Eqs.~(\ref{Eq:measure-of-relativity},~\ref{Eq:rel-VIII}), and 
integrating over transverse momenta, we obtain (this relation 
holds in its unintegrated form)
\be\label{Eq:new-interesting}
    g_1^q(x)-h_1^q(x) = 
    g_T^q(x)-h_L^q(x) \,.
\ee
There are several reasons, why this relation is interesting. 

First, it involves only collinear PDFs, which is the only 
relation of such type in bag model. The QCD evolution equation 
for all these functions are different, which shows the limitation 
of this relation: even if for some reason 
(\ref{Eq:new-interesting}) was valid in QCD at a certain 
renormalization scale $\mu_0$, it would break down at any other 
scale  $\mu\neq\mu_0$. 

Second, for the first Mellin moment this relation is valid  
model-independently presuming the validity of the 
Burkardt-Cottingham sum rule and an analog sum rule for 
$h_L^q(x)$ and $h_1^q(x)$. In QCD there are doubts especially 
concerning the validity of  the Burkardt-Cottingham sum rule. 
However, it is valid in many models such as bag 
\cite{Jaffe:1991ra} or chiral quark soliton 
model~\cite{Wakamatsu:2000ex}.

Third, it would be interesting to learn whether 
(\ref{Eq:new-interesting}) is satisfied in nature approximately. 
Also this relation can be tested on the lattice, especially for 
low Mellin moments and in the flavour non-singlet case. Lattice 
QCD calculations for Mellin moments of $g_T^q(x)$ were reported 
in \cite{Gockeler:2000ja}.

Forth, the relation (\ref{Eq:new-interesting}) can be tested in 
models where collinear PDFs were studied. Some results can be 
found in literature. For example, calculations in the bag 
\cite{Jaffe:1991ra,Signal:1997ct} and spectator 
\cite{Jakob:1997wg} model support this relation. Also one 
counter-example is known: the chiral quark-soliton model does not 
support this relation \cite{Wakamatsu:2000ex,Schweitzer:2001sr}. 
The models supporting (\ref{Eq:new-interesting}) have only the 
components in the nucleon wave-function with the quark orbital 
angular momenta up to $L=0,\,1,\, 2$. The chiral quark soliton 
model, which does not support (\ref{Eq:new-interesting}), 
contains {\sl all} quark angular momenta. 

Fifth, an important aspect of model relations is that they 
inspire interpretations. The relation (\ref{Eq:new-interesting}) 
means that the difference between $g_T^q$ and $h_L^q$ is to the 
same extent a 'measure of relativistic effects in the nucleon' as 
the difference between  helicity and transversity. Both these 
differences are related to the transverse moment of pretzelosity, 
see Eqs.~(\ref{Eq:measure-of-relativity},~\ref{Eq:rel-VIII}) and 
\cite{Avakian:2008dz}.

\section{Pretzelosity and quark orbital angular momentum}
\label{Sec-4:pretzelosity-and-OAM}

In quark models, in contrast to gauge theories, one may 
unambiguously define the quark orbital angular momentum operator 
as $\hat{L}_q^i= 
\bar\psi_q\varepsilon^{ikl}\hat{r}^k\hat{p}^l\psi_q$ (for 
clarity the 'hat' indicates a quantum operator). 
In the absence of gauge fields this definition follows uniquely
from identifying that part of the generator of rotations not 
associated with the intrinsic quark spin. We introduce a 
'non-local version' of this operator 
$\hat{L}_q^i(0,z) = 
\bar\psi_q(0)\varepsilon^{ikl}\hat{r}^k\hat{p}^l\psi_q(z)$
with 
$\hat{r}^k=i\,\frac{\partial\;}{\partial p^k}$ and 
$\hat{p}^l=p^l$
in momentum space. Next let us define the quantity
\be\label{Eq:OAM}
L_q^i(x,p_T)=\int\frac{\di z^-\di^2\vec{z}_T}{(2\pi)^3}\; 
e^{ipz}\;\la 
N(P,S^3)|\bar\psi_q(0)\varepsilon^{ikl}\hat{r}^k\hat{p}^l\psi_q(z)|N(P,S^3)\ra 
\biggl|_{z^+=0,\,p^+ = xP^+} \;.
    \ee
We consider a longitudinally polarized nucleon with the 
polarization vector $S=(0,0,1)$, and focus on the $j=3$ component 
in (\ref{Eq:OAM}). Evaluating (\ref{Eq:OAM}) in the bag model we 
obtain as in \cite{She:2009jq,Efremov:2010cy}
\be\label{Eq:OAM-vs-pretzelosity}
    L_q^3(x,p_T) = (-\,1)\,h_{1T}^{\perp(1)q}(x,p_T)\;.
\ee
Thus, $L^3_q=\int\di x\int\di^2\vec{p}_T L_q^3(x,p_T)= 
(-\,1)\,\int\di x\;h_{1T}^{\perp(1)q}(x)$ is the contribution to 
the nucleon spin from the quark orbital angular momentum of the 
flavour $q$. Adding up the contribution of the intrinsic quark 
spin,  $S_q^3=\frac12\int\di x g_1^q(x)$, we obtain 
$2J_q^3=2S_q^3+2L_q^3=P_q$ which is the consistent result for the 
contribution of flavor $q$ the nucleon spin in SU(6). We stress 
that the relation of  pretzelosity and orbital angular momentum, 
Eq.~(\ref{Eq:OAM-vs-pretzelosity}), is at the level of matrix 
elements of operators, and there is no a priori operator identity 
which would make such a connection.
   
\section{Conclusions}
\label{Sec-6:conclusions}

We presented a study of a complete set relations among T-even 
twist-2 and twist-3 TMDs in the MIT bag model, and discussed to 
what extent these relations are supported in other quark models. 
Special attention was paid to the relation of the difference of 
$g_1^q$ and $h_1^q$ to the (1)-moment of pretzelosity, and the 
relation of the latter to quark orbital angular momentum. It is 
interesting to ask, whether a quark model relation of the type 
(\ref{Eq:OAM-vs-pretzelosity}) may inspire a way to establish a 
rigorous connection between TMDs and OAM in QCD? We hope our 
results will stimulate further studies in quark models.  

Sorrily space limitations did not allow us to address other 
interesting questions like  Lorentz invariance relations 
\cite{Metz:2008ib}, inequalities \cite{Bacchetta:1999kz}, 
Wandzura-Wilczek-type approximations \cite{Avakian:2007mv}, and 
numerical results which support the Gauss Ansatz 
\cite{Schweitzer:2010tt}. All these issues can be found in 
Ref.~\cite{Avakian:2010br}.

\paragraph{Acknowledgements.} {\small 
A.~E.~is supported by the Grants RFBR 09-02-01149 and by the 
Heisenberg-Landau Program of JINR. The work was supported in part 
by DOE contract DE-AC05-06OR23177, under which Jefferson Science 
Associates, LLC,  operates the Jefferson Lab. }

\end{document}